\documentclass[fleqn, 11pt]{wlscirep} % SciRep template
\usepackage[utf8]{inputenc}
\usepackage[T1]{fontenc}

\usepackage{graphics}
\usepackage{xcolor}
\usepackage[normalem]{ulem}
\usepackage{soul}
\usepackage{float}
\usepackage{hyperref}
\usepackage{amsfonts}
\usepackage{amssymb}
\usepackage{lineno}
\DeclareMathAlphabet{\mathcal}{OMS}{cmsy}{m}{n}
\newcommand{\diff}[1]{\mathrm{d}#1\,}

\begin{document}

\title{Investigating structural and functional aspects of the brain's criticality in stroke}

\author[1,+]{Jakub Janarek}
\author[1,+]{Zbigniew Drogosz}
\author[1,2,+]{Jacek Grela} 
\author[1,2,*]{Jeremi K. Ochab} 
\author[1,2,3]{Paweł Oświęcimka}

\affil[1]{Institute of Theoretical Physics, Jagiellonian University, 30-348 Kraków, Poland}
\affil[2]{Mark Kac Center for Complex Systems Research, Jagiellonian University, 30-348 Kraków, Poland}
\affil[3]{Complex Systems Theory Department, Institute of Nuclear Physics, Polish Academy of Sciences, 31-342 Kraków, Poland}

\affil[*]{jeremi.ochab@uj.edu.pl}
\affil[+]{These authors contributed equally to this work.}


\begin{abstract}
This paper addresses the question of the brain's critical dynamics after an injury such as a stroke. It is hypothesized that the healthy brain operates near a phase transition (critical point), which provides optimal conditions for information transmission and responses to inputs. If structural damage could cause the critical point to disappear and thus make self-organized criticality unachievable, it would offer the theoretical explanation for the post-stroke impairment of brain function. In our contribution, however, we demonstrate using network models of the brain, that the dynamics remain critical even after a stroke. In cases where the average size of the second-largest cluster of active nodes, which is one of the commonly used indicators of criticality, shows an anomalous behavior, it results from the loss of integrity of the network, quantifiable within graph theory, and not from genuine non-critical dynamics. We propose a new simple model of an artificial stroke that explains this anomaly. The proposed interpretation of the results is confirmed by an analysis of real connectomes acquired from post-stroke patients and a control group. The results presented refer to neurobiological data; however, the conclusions reached apply to a broad class of complex systems that admit a critical state.

%This paper addresses the problem of the brain's critical behavior in the case of a brain injury such as a stroke.
%\DIFadd{The hypothesis that the healthy brain operates in the regime of critical dynamics, supported by some theoretical and experimental evidence, allows us to explain optimal information processing and its storage in neural networks. However, critical properties of the brain with structural damage are still the subject of a vital discussion. In this contribution, we indicate that even a damaged connectome can be considered critical when statistical measures of criticality are utilized and interpreted with particular care. In particular, we propose a simple model of artificial stroke} to demonstrate that an anomalous behavior of the critical characteristics, the second-largest cluster size, results from the loss of integrity of the network, quantified within the graph theory, and not from genuine non-critical behavior. Thus, even in a post-stroke state, the brain dynamics remain critical. The proposed interpretation of the results is confirmed with the analysis of the real connectomes acquired from post-stroke patients and the control group. The results presented refer to neurobiological data; however,  the conclusions reached apply to a broad class of complex systems for which a critical state is identified.

\end{abstract}

\maketitle

\section{Introduction}

The concept of complexity is used to characterize natural systems consisting of large numbers of nonlinearly interacting elements, resulting in the spontaneous collective behavior of the system on the macroscopic level, called emergence. However, complex systems reveal further intriguing properties; among others, one can mention scale invariance\cite{stanley1999}, self-organized criticality\cite{bak1987,bak1996How,turcotte1999}, and adaptability to new conditions\cite{haken1978}. The variety of complex characteristics makes it impossible to describe the systems by a reductionist approach, i.e., to derive system properties as a simple consequence of a physical law. Characterizing the system structure and dynamics requires rather a holistic approach relying on describing its properties on different levels of organization. In this respect, when the exact mathematical description is unattainable, agent-based modeling\cite{bonabeau2002} is especially beneficial.
Simulating the system as a collection of autonomous entities allows us to explore its dynamics and helps us provide its natural description, which includes emergent phenomena.
%, which allows us to explore the dynamics of the system.
% Simulation of the system as a collection of autonomous entities can help us provide its natural description, which includes emergent phenomena.
This interdisciplinary approach has been applied to study complex systems, encompassing all scientific disciplines, such as physics, chemistry, biology, and social and economic systems\cite{kwapien2012}.  

A canonical example of a complex system is the human brain, whose large numbers of neuronal cells display nontrivial multiscale organization\cite{sporns2004organization,meunier2010modular} %,sporns2010networks ?
and complex characteristics.\cite{chialvo2010emergent,wilting2019} It has also been discovered that power-law statistics, often used to describe critical phase transitions, are present in the brain. These power laws quantify the scale-free properties of neural avalanche distributions, determine the temporal organization of the brain signals recorded from various brain imaging techniques, and characterize the dependence of the correlation length with system size.
The critical brain hypothesis\cite{bak1996How} states that neural networks evolve towards and stay most of their time\cite{tagliazucchi2012,moretti2013NCGriffiths} in a state around a critical phase transition (we will also use interchangeably phrases `at a critical point', `in a critical state' or `at criticality'), where the competition between order and disorder emerges.
Systems in that state have been argued to exhibit optimal computational properties related to information processing, such as information transmission and storage\cite{haldeman2005critical,beggs2008criticality,shew2011information}, computational power\cite{legenstein2007edge} and maximal sensitivity to stimuli\cite{kinouchi2006optimal,shew2009neuronal}.
% , which is especially appealing in neurosciences\cite{zimmern2020}.

\begin{figure}
    \centering  
    \includegraphics[width=0.7\textwidth]{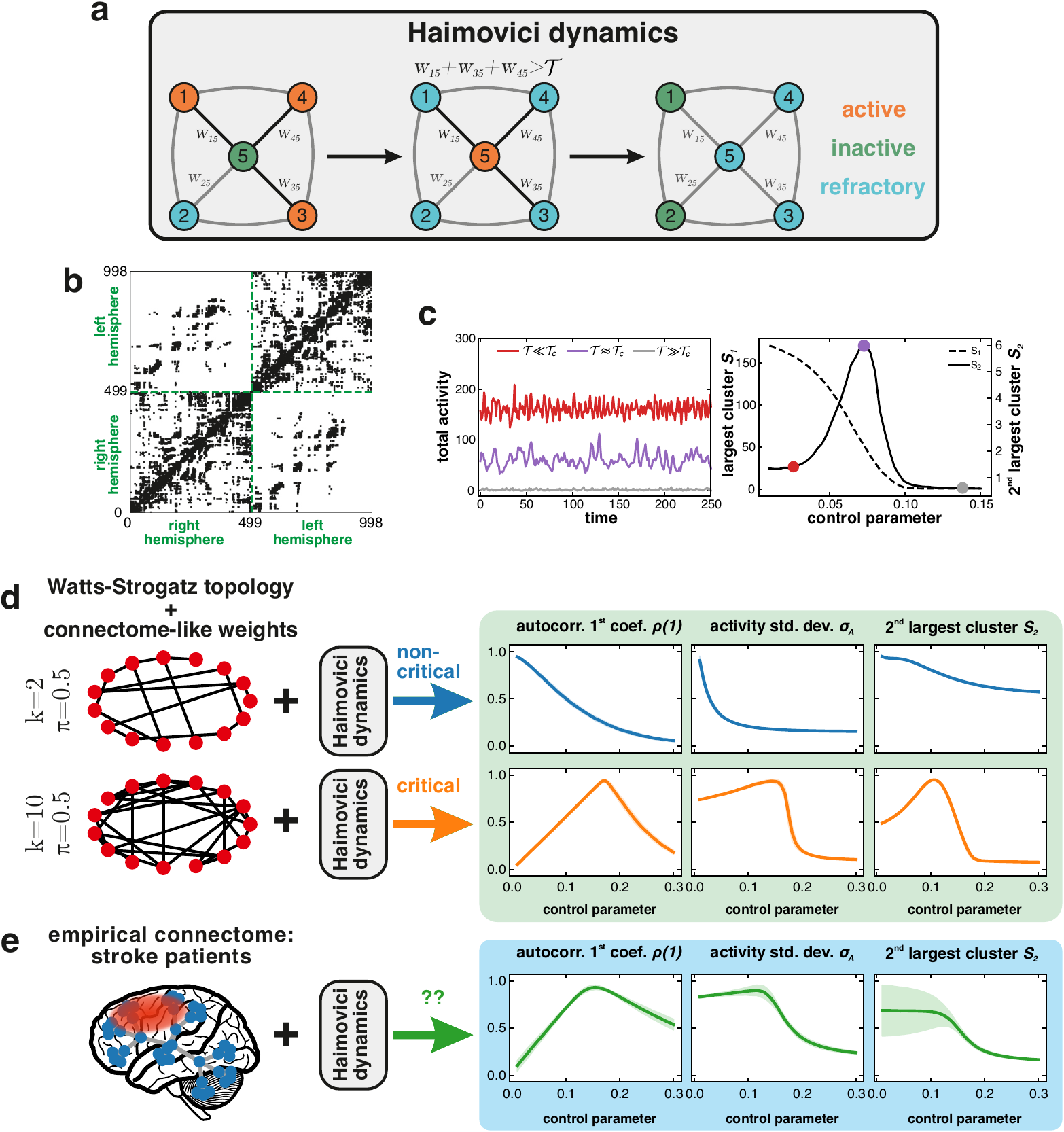}
    \caption{\textbf{Haimovici et al. brain model: description and dependence of criticality status on connectome topology.} \textbf{(a)} Illustration of model dynamics: transitions between possible states of the network nodes (circles with numbers) with connection weights $w_{ij}$. Initially, the inactive (green) central  node is connected to three active (orange) nodes, and to one node in a refractory state (blue). In the next step, assuming that the sum of active neighbors' weights is larger than the threshold, i.e. $w_{15}+w_{35}+w_{45}>\mathcal{T}$, the central node is activated, while the active nodes become refractory.
    In the last step, the transition from the refractory to the inactive state takes place randomly at each node.
    \textbf{(b)} 
    Example of the structural part of the model: Hagmann et al.'s connectome of healthy human subjects, represented by an adjacency matrix. The Haimovici model consists of panel \textbf{(a)} dynamics applied to the connectome-based network. 
    \textbf{(c)} Model criticality for a healthy connectome: total activity and a time-averaged size of the largest ($S_1$) and second-largest ($S_2$) clusters of concurrently active connected nodes, for varying threshold parameter $\mathcal{T}$. For small values of the threshold $\mathcal{T}\ll\mathcal{T}_c$ (red), the nodes are easily activated, while for large values $\mathcal{T}\gg\mathcal{T}_c$ (gray) the nodes remain mostly quiet. Near the critical threshold value $\mathcal{T}\approx\mathcal{T}_c$ (purple), the activity becomes correlated, resulting in a characteristic peak in $S_2$. 
    \textbf{(d-e)} Beyond the Haimovici model: panels show an extension of the model resulting from a replacement of the healthy connectome by artificial networks and post-stroke connectomes.
    The simulated activity is used to compute various indicators of criticality: the time-averaged size of the second-largest cluster $S_2$, the first autocorrelation coefficient $\rho(1)$, and the standard deviation of the total activity $\sigma_A$. \textbf{(d)} Known criticality status: examples of Watts-Strogatz small-world networks resulting in clearly discernible non-critical/critical dynamics (upper/lower part). 
    \textbf{(e)} Stroke patients connectomes: ambiguous criticality status.
    It is readily seen that although both $\rho(1)$ and $\sigma_A$  exhibit local maxima typical  for systems at a critical point, the shape of $S_2$ is less clear-cut and varies considerably between patients (shaded areas denote standard deviations). An explanation of this crucial observation constitutes the central aim of this study.
    \textbf{(d-e)} For the sake of presentation, observables are normalized to their maximal values. For the details of the model, the numerical simulations, and the calculation of observables, see Section~\ref{sec:htc} and Methods~\ref{sec:methods_htc}.
    }
    \label{fig:watts-strogatz_stroke_dataset_comparison}
\end{figure}

% If criticality is indeed a fundamental property of healthy brains2, then neurological dysfunctions shall alter this optimal dynamical configuration. However, we know little about the effect of brain disorders on criticality25.
%\cite{massobrio2015criticality,cocchi2017criticality}
However, the status of criticality and the associated properties in the brain with neurological dysfunction are still not known precisely. 
One possibility is that given critical state provides optimal functioning of the healthy brain, neurological dysfunctions might be associated with its loss\cite{massobrio2015criticality,cocchi2017criticality}, which opens this area of study to clinical applications\cite{zimmern2020}.
% Critical phenomena have also found clinical applications,
These include the study of epilepsy\cite{osorio2010,kramer2012human}, Alzheimer’s and Parkinson’s disease\cite{montez2009, west2016}, and analysis of cognitive processes, including human learning\cite{ouyang2020}.
The ideas of critical phenomena have recently been applied to the study of changes in brain dynamics due to brain damage, both purely computationally\cite{Haimovici2016,goodarzinick2018PRERobustness} and with realistic connectomes of stroke patients\cite{Rocha2022}.
In the latter case, with a simple computational model, the authors were able to predict critical phenomena based on the first principles.
Furthermore, the presence and severity of the stroke reportedly were related to a loss of critical behavior in the brain and a possible post-stroke recovery of a patient to the recovery of the critical state. However, the opposite hypothesis that the brain remains in the critical state even in case of serious injury is also possible.

The problem of assessing whether a system is at a critical point or whether it can exhibit a critical phase transition is even more delicate due to the subtleties of measures of criticality and proper interpretation of the results, and it has been challenging in similar contexts\cite{milanowski2016seizures}.
Calculation of additional criticality-aware quantities, beyond the usual second-largest cluster size, and a comparison with an artificial system with known criticality status
reveals an inconsistency. As summarized in Fig.~\ref{fig:watts-strogatz_stroke_dataset_comparison}, the criticality status of the stroke-affected brain becomes unclear due to an ambiguous behavior of the second-largest cluster size. 
Explaining this crucial observation is our main motivation.

This study is of importance from several perspectives. Brain criticality is an appealing hypothesis implying optimal neural network processing, and a comprehensive understanding of its nature is crucial to understanding brain functioning. However, estimated characteristics of the critical state should be interpreted with particular care since the analysis of the complex systems of which the brain is undoubtedly an example often exhibit nontrivial and subtle properties. Therefore, properties associated with the criticality in the brain with structural damage are still a subject of vital discussion. Moreover, potential deviation from critical neural dynamics could open the possibility for clinical application, as indicated above. In this contribution, we propose a microscopic model of brain stroke to find out the possible mechanism underlying the observed inconsistency in measures associated with neural dynamics at criticality. Such an approach allows us to reproduce the statistics observed in empirical data while controlling the system's inner organization and being able to better characterize it using graph-theoretic tools. Finally, an explanation of the ambiguous character of the criticality indicators exposes the possible difficulty of utilizing a single measure of criticality and offers a consistent data-driven argument to monitor criticality in stroke patients.

The organization of the paper is as follows; in Section~\ref{sec:htc} we describe the data set used, define the Haimovici model and provide its analysis in the context of the post-stroke brain.
Section~\ref{sec:ising} presents the novel model of artificial stroke with its comprehensive analysis. The results of previous sections are jointly discussed and interpreted in Section~\ref{sec:discussion}. Finally, Section~\ref{sec:summary} offers conclusions. Technical details of the graph connectivity measures and Haimovici and Ising models are presented in Section~\ref{sec:methods}.

\section{Criticality in a model of brain activity}
\label{sec:htc}
\subsection{Haimovici et al. brain model}
\label{sec:2-1}
Some crucial aspects of brain dynamics are well reproduced in the critical regime of the cellular automaton-type Haimovici et al. model\cite{Haimovici2013}, where simple dynamical rules are applied to the network of cells based on empirical connectome.  In this section, we demonstrate the quantitative characteristics of the model and discuss their relation to critical and non-critical states. 

The Haimovici et al. model is a three-state cellular automaton\cite{Greenberg1978} on a connectome encoded as a network with weighted connectivity matrix $W$.
In the context of brain activity, each network node represents a region of interest (ROI) of the brain cortex. 
The model dynamics are discrete.
At each time step, a node is in one of three states: inactive (I), active (A), or refractory (R). The transitions between the three states for the $i$-th node of the network are as follows: (i) $I\to A$ always if the sum of the weights of the active neighbors of the node is greater than the activation threshold parameter $\mathcal{T}$, i.e., $\sum_{j  \  \text{active}} w_{ij} > \mathcal{T}$, and with probability $r_1$ otherwise; (ii) $A\to R$ with probability $1$; (iii) $R\to I$ with probability $r_2$.
Figure~\ref{fig:watts-strogatz_stroke_dataset_comparison}\textbf{a} shows an illustrative example of these transitions.
Probabilities $r_1$ and $r_2$ are small numbers, $r_1, r_2\ll 1$, chosen before the start of a simulation, and determine the timescale of the system.
Brain simulations based on the model are summarized in Fig.~\ref{fig:watts-strogatz_stroke_dataset_comparison} and in Methods~\ref{sec:methods_htc}.

% \paragraph{Connectome matrices}
The model dynamics exhibit diverse  behaviors depending on the choice of the connectivity matrix $W$. The authors of\cite{Haimovici2013} used Hagmann et al.'s empirical connectome\cite{Hagmann2008} to find dynamical phase transitions in a healthy brain.
The use of small-world Watts-Strogatz (WS) topology with connection weights mimicking the ones found in empirical connectomes, investigated in\cite{Zarepour2019,Diaz2021}, showed that depending on the network parameters, the model may find itself in different regimes, see Fig.~\ref{fig:watts-strogatz_stroke_dataset_comparison}\textbf{d}, including transience to a ground state, continuous and discontinuous dynamical phase transitions (whereby we mean dis/continuous change of the order parameter, e.g., mean neural activity, as seen in Fig.~\ref{fig:watts-strogatz_stroke_dataset_comparison}\textbf{c}), which all have distinct properties known from physical systems (e.g., in discontinuous transitions hysteresis can be observed).
% Another recent paper\cite{Rocha2022} performed a large-scale study of empirical connectomes of stroke patients focusing on changes in dynamics following a brain lesion and a recovery from it. 

\paragraph{Brain criticality}
% \label{subsec:braincriticality}

The activation threshold $\mathcal{T}$ is the parameter used to control the dynamics of the system. With Hagmann et al.'s connectome\cite{Hagmann2008} as the underlying network, the model admits a dynamical phase transition, and the critical value of the threshold is $\mathcal{T}_c \approx 0.073$ (cf. Methods~\ref{sec:methods_htc}). 
For very small values of $\mathcal{T}$, even the weakest connections between the nodes are enough to spread the activity (supercriticality). Conversely, for large values of $\mathcal{T}$, active nodes may fail to activate their neighbors, and the total activity remains very low (subcriticality; see Fig.~\ref{fig:watts-strogatz_stroke_dataset_comparison}\textbf{c}). It is in the critical regime that the brain activity simulated by the model reproduces the correlations associated with the resting-state networks (RSNs) of the brain\cite{Haimovici2013}. 

The state-of-the-art analysis of model dynamics has been based on the time-averaged sizes of the largest clusters (cf. Methods~\ref{sec:methods_clusters}).
The order parameter can be identified with the average largest cluster size $S_1$, and the 
critical point $\mathcal{T}_c$ is located near the 
local maximum of the size of the second largest cluster $S_2$ as a function of $\mathcal{T}$
(if the maximum exists). 
Such behavior is a strong indicator of a dynamical phase transition\cite{Haimovici2013, Zarepour2019}. The phase transition can also be revealed using other quantities\cite{Haimovici2016}, e.g., the variability of the total activity $\sigma_A$, the variance of the largest cluster size, the first coefficient of the autocorrelation function $\rho(1)$\cite{scheffer2009early,dakos2012robustness}, or the eigenvalues of the correlation matrix.

Figure~\ref{fig:watts-strogatz_stroke_dataset_comparison}\textbf{d} presents three quantities for non-critical (exhibiting only transient dynamics) and critical (exhibiting a continuous phase transition) systems obtained for the Haimovici dynamics on human-connectome-based WS networks. 
The quantities shown are the time-averaged size of the second-largest cluster $S_2$,
the first autocorrelation coefficient $\rho(1)$, and the standard deviation of the total activity $\sigma_A$.
Evidently, each of them has a distinct functional form that allows distinguishing between the non-critical and the critical case.
In Figure~\ref{fig:watts-strogatz_stroke_dataset_comparison}\textbf{e}, the same quantities are calculated for the Haimovici dynamics run on a set of empirical brain connectomes of stroke patients studied in\cite{Rocha2022}. In this case, not all signatures of a critical phase transition are equally clear: $\rho(1)$ and  $\sigma_A$ exhibit a form similar to the system that has a critical point, but $S_2$ behaves in a way that does not correspond clearly either to the non-critical or the critical WS case, in a similar manner to how it was reported in\cite{Haimovici2016}.
An extended investigation of several other quantities, described in Supplementary Information~1, %~\ref{app:otherq}
revealed that the second-largest cluster size was the only observable with ambiguous outcomes.

\subsection{Real and artificial strokes}
\label{sec:real_and_artificial}
The empirical data on strokes were recently studied\cite{Rocha2022} with the use of a dynamical model of brain activity very similar to the one presented in Sec.~\ref{sec:2-1} (cf. Methods~\ref{sec:methods_htc} for details).
The main finding was the observation that in model simulations on connectomes from patients three months after a stroke the maximum in the average size of the second-largest cluster of active nodes
was missing, and it reappeared in a subgroup of those patients when connectomes were acquired again twelve months after the stroke.
The loss and reappearance of the peak were interpreted, respectively, as a loss and recovery of the brain's ability to reach the critical state, parallel to the behavioral post-stroke recovery of a patient.

% We shall briefly describe the stroke data set used, with an emphasis on the usage of the Haimovici model (introduced above).
In order to explain the origin of these findings, 
we propose a minimal model of artificial strokes that recreates two key features found in the empirical data: a) the signatures of criticality summarized in Fig.~\ref{fig:watts-strogatz_stroke_dataset_comparison}\textbf{e}, and in particular the anomalous behavior (the absence of a peak) in the second-largest cluster size, and b) the decrease in connectome integrity correlated with this behavior. 
%We argue that they are sufficient to explain the confusing character of criticality-aware observables. 

% A recent paper\cite{Rocha2022} studied the behavior of clusters in a model of brain activity in individual connectomes acquired from stroke patients and the control group via Diffusion-Weighted Magnetic Resonance Imaging (DWI). 
% The main finding of the paper was the observation that in simulations of the model on connectomes from patients three months after a stroke the maximum in the average size of the second-largest cluster of active nodes
% was missing, and it reappeared in a subgroup of those patients when connectomes were acquired again twelve months after the stroke.
% The loss and reappearance of the peak were interpreted, respectively, as a loss and recovery of the criticality of the brain, parallel to the behavioral post-stroke recovery of a patient.
% The model used to study the dynamics of brain activity in the paper cited was very similar to the one presented in Sec.~\ref{sec:htc} (cf. Methods~\ref{sec:methods_htc} for details).

\paragraph{A minimal model of artificial strokes}

We introduce a model of a stroke-like modification to a healthy connectome.
% The Hagmann et al.'s connectome\cite{Hagmann2008} serves as the base connectivity matrix representing the healthy brain.
Given a healthy empirical connectome, an artificial stroke changes the connectivity between a particular RSN with the rest of the brain.
To this end, we randomly select a fixed fraction of nodes (a proxy of stroke severity) of the RSN and completely remove connections to their neighbors not belonging to the same RSN.
This way, the internal structure of the RSN remains unchanged, while effectively decreasing the connection of the RSN with the rest of the brain.
The model we propose aims to reproduce global characteristics found in real connectomes affected by a stroke and does not purport to offer a biologically or neurologically plausible mechanism.

\begin{figure}[h]
    \centering
    \includegraphics[width=0.9\textwidth]{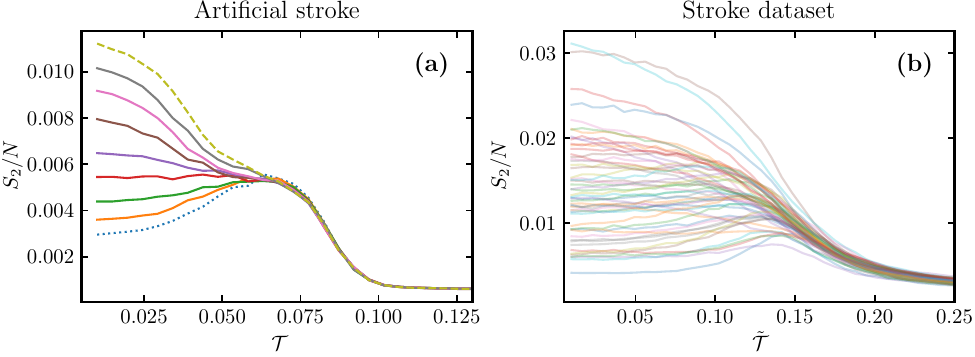}
    \caption{
    \textbf{Size of the second largest cluster $S_2$ as a function of the threshold parameter $\mathcal{T}$ for artificial and real stroke.} \textbf{(a)} Results for an artificial stroke afflicting the auditory RSN with varying stroke severity.
    The color lines correspond to the fraction of nodes in the RSN disconnected from the rest of the brain, ranging from $75\%$ (the bottom-most dotted curve) to all (the topmost dashed curve).
    \textbf{(b)} Results for connectomes from the stroke dataset in both control and patient groups (cf. Fig 3b in Rocha et al.\cite{Rocha2022}). Each line corresponds to one person. \textbf{(a-b)} Artificial strokes successfully recreate an anomalous loss of peak in the $S_2$ curve.
    }
    \label{fig:rsn3_and_strokedataset}
\end{figure}

In Figure~\ref{fig:rsn3_and_strokedataset}, we compare the second-largest cluster size  calculated for the Haimovici dynamics on (a) connectomes with an artificial stroke of increasing severity located in a single RSN and (b) real post-stroke connectomes.
% \cite{Rocha2022}.
In the latter case, the connectivity matrices were normalized (see Methods~\ref{sec:methods_htc}) thereby shifting the critical point $\mathcal{T}_c \rightarrow \tilde{\mathcal{T}}_c$.
We report a good qualitative agreement between the outputs of the proposed artificial stroke model and the real stroke dataset, as in both models the second-largest cluster sizes deviate from the standard (i.e. healthy) case with a pronounced peak.

\paragraph{Structural analysis}

\begin{figure}[ht!]
    \centering
    \includegraphics[width=0.9\textwidth]{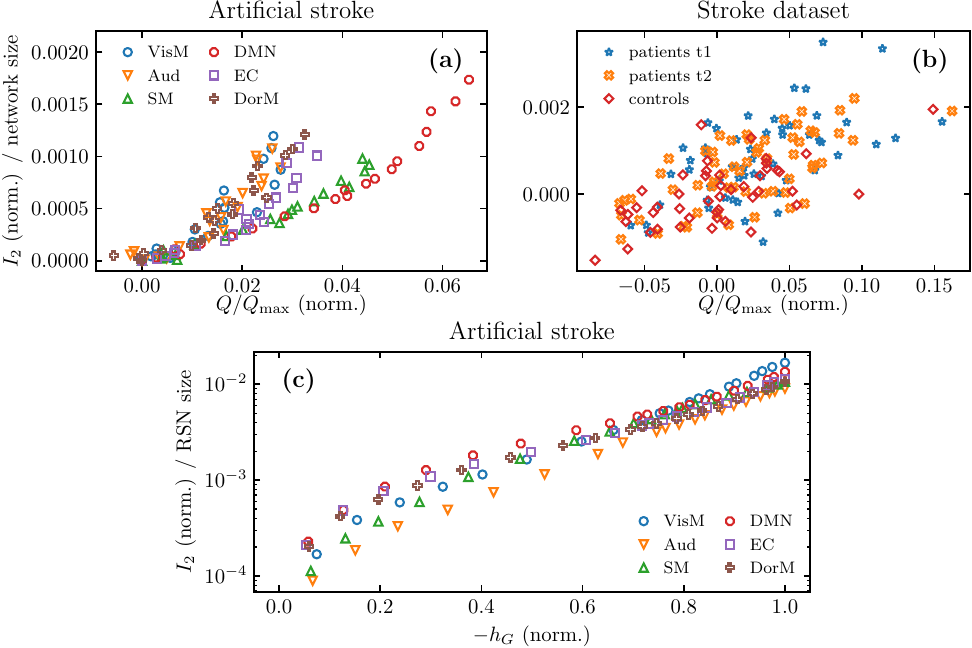}
    \caption{ 
    \textbf{Structural analysis of stroke-affected connectomes.} 
    \textbf{(a)} Normalized area under the $S_2$ plot versus normalized modularity for the artificial strokes on Hagmann et al.'s connectomes with single RSNs gradually more disconnected from the rest of the brain, starting from $0\%$ to $100\%$ of nodes (from zero to high normalized modularity). The Pearson's correlation coefficient between the normalized area under $S_2$ and the normalized modularity is $\rho = 0.88$, p-value $\ll 0.001$ with 95\%CI$[0.84, 0.92]$.
    \textbf{(b)} Similar to \textbf{(a)} but for empirical connectomes from the real stroke dataset for three patient groups with correlation coefficient $\rho = 0.646$, p-value $\ll 0.001$, 95\%CI$[0.55, 0.73]$.  
    \textbf{(a--b)} Both types of strokes with variable severity result in an increase in the area under the $S_2$ curve as well as the values of modularity. The latter is related to the loss of subsystem interconnectivity of the stroke-affected connectome. 
    \textbf{(c)} Normalized area under the $S_2$ plot for the Haimovici model versus the normalized conductance for Hagmann et al.'s connectomes with an increasing fraction of RSN's nodes disconnected from the rest of the brain, from $0\%$ to $100\%$ (from low $-h_G \text{(norm.)}$ values to $-h_G \text{(norm.)} = 1$). The correlation is $\rho = 0.88$ with p-value $\ll0.001$, 95\%CI$[0.84, 0.92]$.
    }
    \label{fig:structural_analysis}
\end{figure}

We provide a deeper analysis of the dynamics of the Haimovici model and connectome integrity using graph-theoretic methods. In Figure~\ref{fig:structural_analysis}, the dynamical part is summarized on the y-axis by the area under the $S_2(\mathcal{T})$ curve $I_2 = \int S_2(\mathcal{T})\diff{\mathcal{T}}$, which qualitatively captures the loss of a peak, explained by high values of $S_2$ for low $\mathcal{T} \leq \mathcal{T}_c$. The connectome integrity is probed by the normalized modularity $Q$ where the modules are found using the Louvain algorithm (cf. Methods~\ref{sec:methods_graph} for details).
In Figure~\ref{fig:structural_analysis}\textbf{a} we present results for artificial strokes with six affected RSNs and different severity levels, i.e., disconnection of between $0\%$ and $100\%$ of the RSN nodes from the rest of the brain, while in Fig.~\ref{fig:structural_analysis}\textbf{b} we plot real stroke results for three patient groups. The normalization is such that mildly affected dynamics are centered near the $(0,0)$ point, while dynamics severely affected by a stroke tend to move away from the origin.

We provide an additional analysis of the artificial stroke in terms of the conductance, $h_G$ (cf. Methods~\ref{sec:methods_graph}), an alternative measure of connectivity between network subsystems\cite{Chung1997}.
This measure, unlike modularity, quantifies the connectivity between a selected RSN and the rest of the connectome.
It offers an additional test of the loss of  integrity between known regions rather than any changes due to Louvain algorithm reconfiguring modules after the stroke. Figure~\ref{fig:structural_analysis}\textbf{c} presents a plot of the normalized area $I_2$ versus normalized conductance $h_G$. We reveal a strong correlation between the network integrity and the anomalous behavior of the second-largest cluster size.

\section{Clusters in a divided Ising model}
\label{sec:ising}

In the previous section, we established that the loss of connectome integrity coincides with the anomalous behavior of the second-largest cluster size in both artificial and real strokes.
In light of this relation, we continue to investigate the changes in connectome structure to understand the development of this anomalous behavior in more detail and reintroduce the question of criticality: is the brain's critical state reachable, or is it lost after the stroke?
In this part, we tackle it by combining the insights from connectome integrity with the Ising model, a paradigmatic case with an existing critical phase transition. Ising spins act as the neuron nodes, and the usual two-dimensional grid takes the role of the connectome. The stroke-induced loss of integrity is taken to an edge case where the connectome is completely divided into subsystems, thus modeling a severe artificial stroke. In particular, we recreate an anomalous lack-of-peak in the second-largest cluster size and describe the underlying mechanism as a competition within the hierarchy of subsystem-wide clusters.

\subsection{Model description}
The Ising model is among the simplest systems that undergo a continuous phase transition as the temperature $T$ changes\cite{Onsager1944,McCoy2014,fraiman2009,camia2009}. 
In a similar context, it has been used to study and reproduce the behavior of neuronal populations in cultured cortical neurons, cortical slices, and visual cortex among others\cite{tang2008maximum, yu2008small,mora2011biological}, and in particular to study structural damage to functional networks at criticality\cite{goodarzinick2018PRERobustness}. Another study compared the fMRI correlations with the correlation matrices obtained from Ising model simulated on empirical connectomes to characterize disorders of consciousness with the model's critical temperature.\cite{abeyasinghe2020JCMConsciousness}
The Ising model itself consists of a network whose sites take the values $\pm 1$ (originally representing two states of atomic spins) together with a particular definition of interactions between adjacent spins, which describes the time evolution of spin states (cf. Methods~\ref{sec:methods_ising}).

The usual indicator of a continuous phase transition is the correlation length, i.e., the characteristic length-scale below which the non-adjacent spins are correlated, and which diverges at the transition point.
In addition, it is agreed upon that observables related to clusters, used in percolation theory, are viable probes of critical state in the two-dimensional Ising model\cite{Coniglio1977}.
Clusters (domains) are defined as maximal connected sets of sites with the same orientation of the spins. The size of a cluster is the number of its sites.
The size distribution of the clusters depends crucially on $T$ and is used to construct two well-established indicators of criticality: (a) an abrupt change in the time-averaged size of the largest cluster $S_1$ and (b) a peak in the time-averaged size of the second largest cluster $S_2$; both occur near the critical temperature $T_c$\cite{CZLLW2021}.

\begin{figure}[H]
    \centering
    \includegraphics[width=0.65\textwidth]{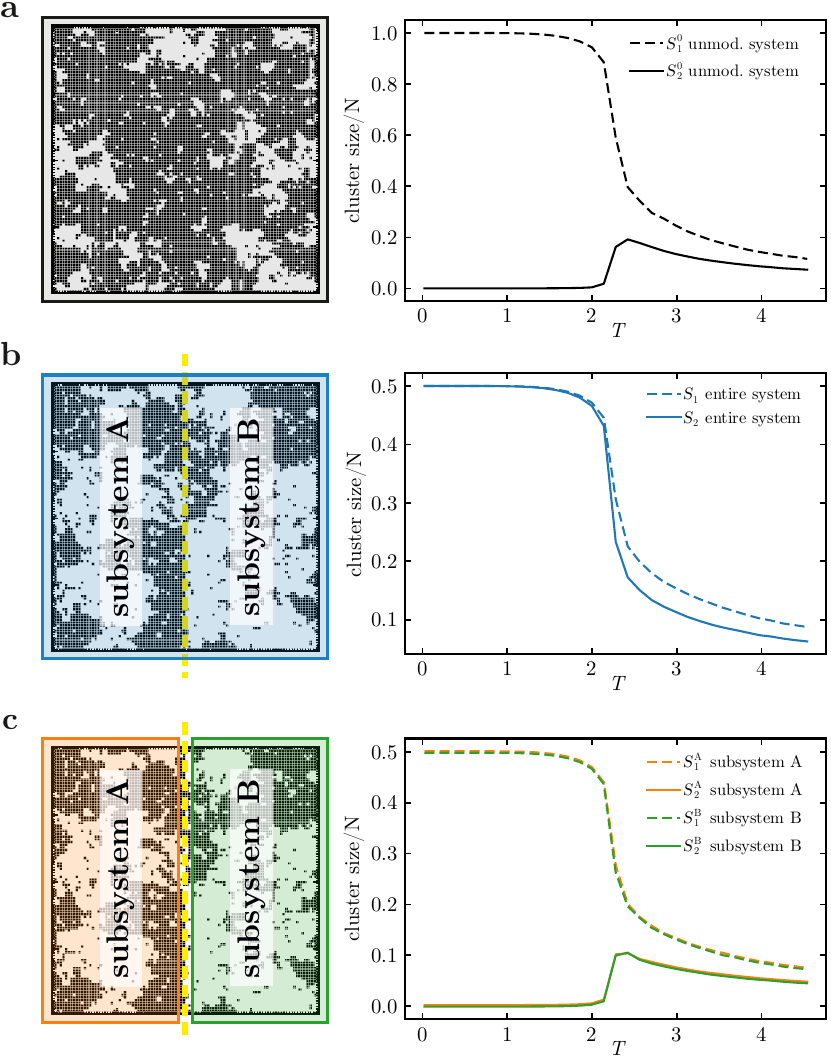}
    \caption{
    \textbf{Divided Ising model: temperature dependence of two largest cluster sizes.}
    A schematic representation of a square lattice and dependence of average  sizes of the largest cluster $S_1$ (dashed lines) and of the second-largest cluster $S_2$ (solid lines) on temperature $T$. \textbf{(a)} In an undivided lattice (upper index $0$), clusters exhibit a characteristic saturation of $S_1^0$ for small temperature and a maximum of $S^0_2$ in the vicinity of the critical temperature $T_c \approx 2.27$. Panels \textbf{(b-c)} show a system divided into two equally sized subsystems (as marked by the yellow dashed line on the lattices). Panel \textbf{(b)} shows the largest ($S_1$) and second-largest ($S_2$) clusters of the entire system. The graphs of $S_1$ and $S_2$ have both qualitatively \emph{the same shape}, saturating for small $T$ at half of the system size. The temperature dependence of both is similar to that of the largest cluster of the unmodified system ($S_1^0$ in panel \textbf{(a)}), and in $S_2$ no peak is present. Panel \textbf{(c)} shows the largest and second-largest clusters computed for each of the subsystems separately (superscripts A/B, green and orange lines). The curves are down-scaled versions of the clusters for the unmodified system (panel \textbf{(a)}; note the different scale of the $y$-axis), with the signature peak in $S_2$. Corresponding curves for the subsystems fully overlap.
    }
    \label{fig:ising_clusters}
\end{figure}

\begin{figure}[H]
    \centering
    \includegraphics[width=0.9\textwidth]{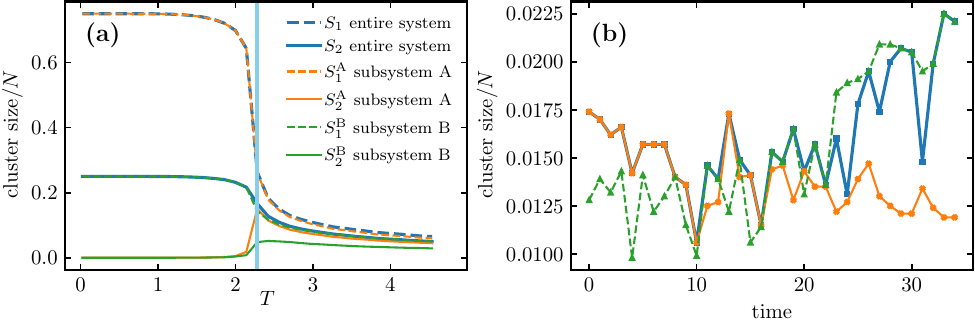}
    \caption{
    \textbf{Dependence of cluster sizes on temperature and time in the two-dimensional Ising model divided into two fully disconnected subsystems of unequal size.} Subsystem A consists of $N_\text{A}=7500$ sites and subsystem B of $N_\text{B}=2500$ sites; see Fig.~\ref{fig:ising_patches} for the geometric arrangement. \textbf{(a)} Temperature dependence of the largest and second-largest cluster sizes for the entire system (blue lines), subsystem A (orange lines), and subsystem B (green lines). The line styles follow Fig.~\ref{fig:ising_clusters}.
    In subsystems, characteristic features such as saturation of $S_1$ at small temperatures and a peak in $S_2$ near the critical $T_c$ persist, but the size symmetry breaking with $N_\text{A} > N_\text{B}$ results in relative growth of clusters related to subsystem A with respect to subsystem B. The vertical blue line is a near-critical region where we probe the time evolution to be shown in the right panel. \textbf{(b)} Interplay between momentary clusters $S_1^\text{B}(t)$, $S_2^\text{A}(t)$ and $S_2(t)$ over a number of time steps at a fixed $T$ (the line styles are the same as for the averages in panel \textbf{(a)}). One can observe the competition for the place of the second largest cluster (blue solid line), with $S_2^\text{A}(t)$ (orange solid line) dominating earlier and $S_1^\text{B}(t)$ (green dashed line) later.
    } 
    \label{fig:ising_patch_50x50}
\end{figure}

We introduce a key modification to the model's connectivity matrix by removing certain links from the original two-dimensional square lattice to form two completely disjoint subsystems, A and B. This mimics the loss of integrity, studied previously in the case of strokes, whereas the Ising dynamics fixes the critical behavior. We consider two cases: subsystems of (a) equal $N_\text{A} = N_\text{B}$, Fig.~\ref{fig:ising_clusters}, and (b) unequal $N_\text{A} > N_\text{B}$ number of nodes, Fig.~\ref{fig:ising_patch_50x50}. In what follows, we inspect the average cluster sizes $S_1$ and $S_2$ as indicators of criticality in the entire system (no superscript) and in each subsystem separately (superscripts A and B). Details of the simulations are discussed in Methods~\ref{sec:methods_ising}.

\paragraph{Inconsistent indicators of criticality} 

In Figure~\ref{fig:ising_clusters}a, we revisit the average sizes of the two largest clusters in the standard Ising model on a $100 \times 100$ square lattice. The size of the largest cluster (black dashed line) saturates at the size $N$ of the entire system; the size of the second-largest cluster (black solid line) starts at zero for small $T$, increases to reach a peak near the critical temperature $T_c$, and decreases to a certain nonzero value for large $T$ (a finite-size effect; $S_2/N \rightarrow 0$ for $N  \rightarrow \infty$). 

In the next step shown in Fig.~\ref{fig:ising_clusters}\textbf{b}, we consider the divided Ising model with two fully disconnected subsystems of equal size and shape (rectangles $100 \times 50$). In this case, cluster sizes $S_1$ and $S_2$ both admit a similar functional form, as shown by the dashed and solid blue lines. Both quantities saturate similarly to the largest cluster calculated for the undivided system and have no maximum near $T_c$. 

Via an introduction of a simple division of the system into two non-interacting parts we recreate the anomalous lack of a peak in the second-largest cluster size $S_2$. To gain insight into how this emerges, we plot in Fig.~\ref{fig:ising_clusters}\textbf{c} the average sizes of the two largest clusters but restricted to each subsystem: $S_1^\text{A}, S_2^\text{A}$ for subsystem A (green) and $S_1^\text{B}, S_2^\text{B}$ for subsystem B (orange). Unsurprisingly, the respective curves for each subsystem overlap and their functional form is identical, up to a rescaling, as that of the cluster size curves for the undivided lattice, described above and shown in the inset of Fig.~\ref{fig:ising_clusters}\textbf{b}. Since the two subsystems are fully disconnected, each has its own independent Ising dynamics: Both are critical around $T_c$, and their respective cluster sizes show typical signatures of criticality. However, the lack of the characteristic peak in the size of the second-largest cluster of the entire system shown in Fig.~\ref{fig:ising_clusters}\textbf{b} is clearly flawed by suggesting the absence of a critical phase transition in the divided system.
% We believe this observation lies at the source of the confusion about  criticality illustrated in Fig.~\ref{fig:watts-strogatz_stroke_dataset_comparison}
% and  resolving it requires 

\paragraph{Cluster ordering}
Below we refine our understanding of the \emph{ordering} of entire-system clusters in terms of subsystem clusters.
As long as the subsystems are fully disconnected and thus have completely separate dynamics within each subsystem, it is the subsystem clusters that play a primary role, and the system-wide clusters only name the largest among them. The lack of a peak in the size of the second largest cluster observed at the level of the entire system comes merely from a particular ordering of the sizes of the subsystem clusters.

At each time step $t$, we are interested in the two largest clusters of the entire system $S_1(t)$, $S_2(t)$ and of each of the subsystems $S_1^\text{A}(t),S_2^\text{A}(t),S_1^\text{B}(t),S_2^\text{B}(t)$. Note that in this paragraph we use momentary cluster sizes, not time averages. We can compare the sizes of the four above-mentioned subsystem clusters and write them down in a list in decreasing order. The first two entries in the list are the largest entire-system clusters: $S_1(t) = \max_1 \left (S_1^\text{A}(t),S_2^\text{A}(t),S_1^\text{B}(t),S_2^\text{B}(t) \right ) = \max \left (S_1^\text{A}(t),S_1^\text{B}(t) \right )$ and $S_2(t) = \max_2\left (S_1^\text{A}(t),S_2^\text{A}(t),S_1^\text{B}(t),S_2^\text{B}(t) \right )$, where $\max_i$ selects the $i$-th largest number. It is clear that regardless of the definitions of the two subsystems, $S_1(t)$ is either $S_1^\text{A}(t)$ or $S_1^\text{B}(t)$, so the functional form of the average $S_1$ corresponds to that of the largest subsystem cluster. Crucially, however, since the role of $S_2(t)$ can be assumed by any of the remaining clusters $S_1^\text{A}(t), S_2^\text{A}(t), S_1^\text{B}(t), S_2^\text{B}(t)$, the functional form of the average $S_2$ depends on the relative size of the subsystem clusters and the robustness of their ordering (i.e., whether the order of the subsystem clusters on the list remains constant throughout the simulation). 

If $N_\text{A}=N_\text{B}$, then $\{S_1(t),S_2(t)\} = \{S_1^\text{A}(t),S_1^\text{B}(t)\}$ for most of the simulation time (where $=$ denotes set equality). In other words, the roles of the two largest entire-system clusters are decided by the competition of the largest cluster of subsystem A and the largest cluster of subsystem B. Therefore, $S_1$ and $S_2$ both share the typical characteristics of the largest subsystem cluster, resulting in the same functional behavior, as shown by the blue lines in Fig.~\ref{fig:ising_clusters}\textbf{b}. In the next section, we demonstrate the results of decreasing the robustness of the ordering by bringing $S_2^\text{A}$ and $S_1^\text{B}$ to comparable sizes.

\paragraph{Competition between clusters}

In the discussion above, we described how the momentary largest system clusters are selected from an ordered list of subsystem clusters. In this section, we 
apply this to the case of unequally sized subsystems $N_\text{A} > N_\text{B}$, where the ordering is less stable in the simulation time. We use a square lattice of the same dimensions as previously but with subsystem B redefined as a smaller square patch in the middle of the lattice and with the connections changed accordingly, as shown in Fig.~\ref{fig:ising_patches}. This reduces the average sizes of the subsystem-B clusters $S_1^\text{B}$ and $S_2^\text{B}$. For a certain patch size, the typical size of the largest cluster in subsystem B $S_1^\text{B}$ is comparable to the second largest cluster in subsystem A $S_2^\text{A}$, and these two clusters compete for the role of the second largest cluster in the entire system $S_2$. 

In Figure~\ref{fig:ising_patch_50x50}, this mechanism is demonstrated on a square lattice of $N=10000$ sites divided into $N_\text{A}=7500$ and $N_\text{B}=2500$ sites. The loss of symmetry between the subsystem sizes results in a different temperature dependence of the cluster sizes. 
In the case of equally sized subsystems shown in Fig.~\ref{fig:ising_clusters}\textbf{b} and described in previous sections, the average cluster sizes in subsystems A and B overlap: $S_1^\text{A} \approx S_1^\text{B}$ (dashed green and orange lines) and $S_2^\text{A} \approx S_2^\text{B}$ (solid green and orange lines). On the other hand, the unequally sized case discussed currently is presented in Fig.~\ref{fig:ising_patch_50x50}a and shows a clear cluster size asymmetry: $S_1^\text{A} \gg S_1^\text{B}$ and $S_2^\text{A} \gg S_2^\text{B}$ near the critical temperature indicated by the blue vertical line. In this region, moreover, $S_2^\text{A} \approx S_1^\text{B}$, which signals the breaking of the cluster ordering.

The competition between the clusters $S_2^\text{A}$ and $S_1^\text{B}$ is shown more closely in Fig.~\ref{fig:ising_patch_50x50}b. In this panel, we plot the dynamics of the clusters of interest at a fixed temperature $T\approx T_c$. 
The solid orange line and the dashed green line trace the temporal evolution of the size of the clusters that compete for the role of the entire-system cluster $S_2$. In this particular time window, we observe a reversal of the roles, in which the role of $S_2$ is played by $S_2^\text{A}$ in earlier time steps and by $S_1^\text{B}$ in later time steps. The plot reveals rich intermittent dynamics that elucidate the averaged picture in Fig.~\ref{fig:ising_patch_50x50}a. The cluster order is perturbed even beyond the competition of $S_2^\text{A}$ and $S_1^\text{B}$, since, at certain moments such as $t=31$, the cluster $S_1^\text{B}$ is temporarily able to take over the role of $S_1$ and reduce the cluster $S_1^\text{A}$ to the role of $S_2$.

% \section{Discussion: Common mechanism for the loss of peak in the Ising and Haimovici models}
\paragraph{Loss of peak: A mechanism common to the models of Ising and Haimovici}

\begin{figure}[H]
    \centering
    \includegraphics[width=0.9\textwidth]{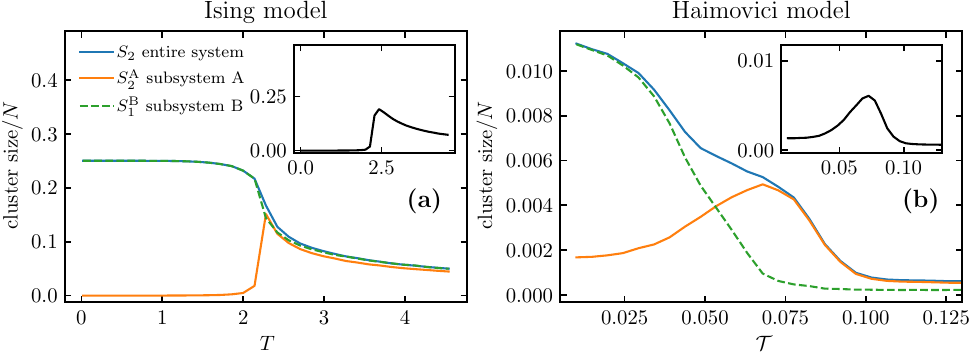}
    \caption{
    \textbf{Cluster sizes in fully disconnected subsystems in the Ising and Haimovici models.} \textbf{(a)} The Ising model with subsystems of unequal size $N_\text{A} = 7500$, $N_\text{B} = 2500$. (This example was shown in Fig.~\ref{fig:ising_patch_50x50}.) The legend is valid for both plots. \textbf{(b)} The Haimovici model with subsystems of sizes $N_\text{A} = 879$ and $N_\text{B} = 119$, where B denotes the auditory RSN, and A is the rest of the network.
    \textbf{(a-b)}
    In the unmodified systems (insets), $S_2$ shows a characteristic maximum near the critical point.
    With subsystem B disconnected, the second largest cluster saturates (panel \textbf{(a)} Ising) or grows monotonically (panel \textbf{(b)} Haimovici) (blue solid line) as we lower $\mathcal{T}$. In both cases, the second largest cluster in subsystem A (orange solid line) exhibits a maximum around the critical point, however, just below this point, near $\mathcal{T}\approx 0.05$ in the Haimovici model ($T \approx 2.2$ in the Ising case) the size of the largest cluster in subsystem B (dashed green line) becomes larger and takes over in the cluster size hierarchy.
    }
    \label{fig:rsn3_disconnected}
\end{figure}

In Fig.~\ref{fig:rsn3_disconnected}, we juxtapose cluster analyses performed on (a) the divided Ising model of unequal sizes and (b) the Haimovici model with severe artificial stroke resulting in completely disconnected auditory RSN (different choices of base RSNs give similar results; see Supplementary Fig.~\ref{fig:supp_rsn_cluster_hierarchy}
in Supplementary Information~\ref{sec:suppA}).
For a detailed description of how the Ising model and the Haimovici brain model are different and how they allow a direct comparison, cf. Methods~\ref{sec:methods_comparison}.
In both models, the average second largest cluster size $S_2$ of the entire system changes its behavior in the same way: When subsystem B — or the auditory RSN — is completely disconnected, $S_2$ loses its characteristic critical peak. The numerical data for both models fully support our prediction: At some value of the threshold parameter $\mathcal{T}$, the size of the largest cluster of the smaller subsystem $S_1^\text{B}$ exceeds the size of the second largest cluster of the larger subsystem $S_2^\text{A}$. This change in the cluster size hierarchy leads to $S_2$ monotonically decreasing as a function of $\mathcal{T}$.

\begin{figure}[H]
    \centering
    \includegraphics[width=0.9\textwidth]{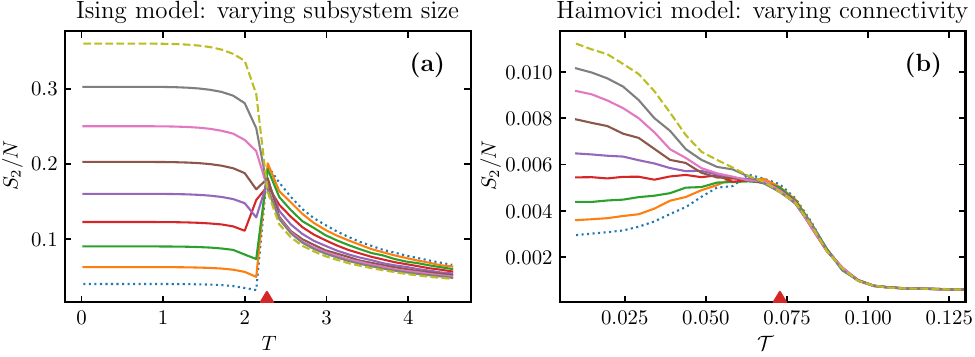}
    \caption{
    \textbf{The size of the second largest cluster in gradually modified systems.} 
    \textbf{(a)} The Ising model with varying subsystem size. The family of color lines represents $S_2$ for different sizes of subsystem B ranging from $N_\text{B}=400$ (the bottom dotted dark blue line) to $N_\text{B}=3600$ (the topmost dashed yellow line). Above a certain size of subsystem B, the critical maximum is lost. The red arrow marks the critical temperature.
    \textbf{(b)} The Haimovici model with changing RSN connectivity. The color lines represent $S_2$ for connectomes with a varying number of connections between the auditory RSN and the rest of the brain removed, ranging from $75\%$ of the connections removed (the bottom-most dotted light blue curve), to all connections between the RSN and the rest of the brain removed (the topmost dashed yellow curve). The red arrow indicates the critical value of the threshold.  \textbf{(a-b)} In both models, above a certain proportion of removed connections, the maximum in $S_2$ is lost.
    }
    \label{fig:ising_htc_comparison}
\end{figure}

In Figure~\ref{fig:rsn3_disconnected}, we presented the edge case of a fully disconnected subsystem, or a severe artificial stroke. In Figure~\ref{fig:ising_htc_comparison}, we extend the presentation to a comparison of gradual system modifications in (a) the Ising model with various sizes of the disconnected subsystem B and (b) the Haimovici model for varying degrees of connectivity between the auditory RSN, chosen as the smaller subsystem, and the rest of the network. Both the variation in the size of the smaller Ising subsystem and changes in connectivity between RSNs produce similar comb-like families of lines (cf. Fig.~\ref{fig:rsn3_and_strokedataset} for similar plots comparing connectomes affected by artificial and real strokes). These modifications are not strictly equivalent, but they provide a clear presentation of the idea. In Supplementary Information~\ref{sec:suppA},
we expand on the present analysis with complementary modifications. The small perturbation regime, where the subsystem has relatively few nodes (in the Ising model) or few interconnections are removed (in the Haimovici model), constitutes the lower part of the comb. In this region, the entire system cluster $S_2$ has typical critical behavior with a peak near the critical $T_c$ or $\mathcal{T}_c$. The large perturbation regime is, in turn, reached when the subsystem is large (the Ising model) or almost completely disconnected (the Haimovici model), and constitutes the upper part of the comb. The upper limiting cases are reached if subsystem B occupies about half of the system or if the auditory RSN is fully disconnected, as studied in Figs.~\ref{fig:ising_clusters}\textbf{b} and~\ref{fig:rsn3_disconnected}, respectively. In this regime, the cluster $S_2$ loses its characteristic criticality peak. 

\section{Discussion}
\label{sec:discussion}
By considering the divided Ising model, we explained the rise of anomalous lack of peak in the size of the second-largest cluster of active nodes, $S_2$, in terms of subsystem clusters forming an ordering for the entire system.
We believe this observation lies at the source of the confusion about  criticality illustrated in Fig.~\ref{fig:watts-strogatz_stroke_dataset_comparison}.
It should be emphasized that each subsystem separately preserves the typical behavior of clusters related to criticality, i.e., growth of the largest cluster, $S_1$, and a maximum in the second largest cluster size, $S_2$.
The critical phase transition is preserved, although it becomes obfuscated by the loss of connectome integrity.
Let us note, that there might be another contributory factor: a shift in $T_c$ connected to the change in the number of nodes in the network, as observed in a related model\cite{deco2012FCNHow}.
This effect, however, does not account for disparate behaviors of the criticality indicators, nor is it strong enough to explain a full loss of phase transition.

We were able to consistently translate the above explanation to the Haimovici model.
There, an apparent loss of criticality was a consequence of changes in the ordering of subsystem clusters enabled by the changes in connectome integrity.
In both models gradual subsystem changes led to the comb-like behavior of $S_2$, Fig.~\ref{fig:ising_htc_comparison}, seen previously in the real stroke data in Fig.~\ref{fig:rsn3_and_strokedataset}\textbf{b}.
Due to the particularities of the Ising dynamics, instead of varying the connectivity, which would render the comb a single line, we varied the subsystem size, which reproduces the expected effect.
We discuss these complementary approaches in Supplementary Information \ref{sec:suppA}.
A similar subsystem-cluster analysis was performed with each of the eight RSNs entirely disconnected, see Supplementary Information, resulting in a similar effect.

Crucially, the change in connectome integrity was the primary driver while the dynamics itself was secondary. %, assuming the critical phase transition was present.
Despite the limiting assumptions, the artificial stroke model recreated quite well the overall loss of connectome integrity in Figure~\ref{fig:structural_analysis}.
In both artificial and real strokes, we found that a stroke-induced loss of connectome integrity measured by an increase of the modularity $Q$ and the decrease of conductance $h_G$ coincides with the anomalous behavior of the second-largest cluster size.

While the authors of another study simulating Ising dynamics on connectomes\cite{abeyasinghe2020JCMConsciousness} hypothesize that structural connectivity alone cannot explain some effects observed in functional correlations of patients after severe brain injuries, no other explanation has been provided in the literature.
Our explanation of the anomalous behavior of the cluster-based indicator of criticality, on the other hand, is consistent with the distinction\cite{martin2020NJPIntermittent,barzon2022JPCCriticalitya} between well-known structurally-driven percolation phase transitions and dynamic transitions—for which less is known.\cite{Diaz2021} Reframed in these terms, our results would suggest that in brain strokes there is no change in whatever underlies dynamic transitions, but that there are structural changes that could disorganize the percolation-like transitions.

\section{Summary and conclusions}\label{sec:summary}
In this study, we revisited the question of whether post-stroke brain dynamics stay at the critical point.
To this end, we compared how indicators of criticality behave after real strokes and computer-simulated strokes proposed for this purpose,
and we do not find evidence for the post-stroke loss of critical dynamics as previously suggested.
Rather, we show that elementary indicators of criticality should be interpreted with caution.
In particular, the behavior of the size of the second-largest cluster of activity in the brain affected by stroke may result solely from the loss of connectome integrity without the brain's departure from a dynamic critical transition.
From this perspective, the behavior is understood in terms of subsystem clusters competing for the top rank system-wide.
The results have been reproduced in classic physical models and confirmed based on graph-theoretical characteristics calculated for the empirical connectomes.
Thus, when a system with an unknown subsystem structure is analyzed, stroke-induced loss of critical dynamics may be illusory, and the simpler and more plausible explanation is the described mechanism cluster competition mechanism enabled by a loss of connectome integrity alone.

\subsection{Outlook}
We hope that as the concept of criticality becomes increasingly relevant\cite{zimmern2020}, and the critical state of the brain is evaluated in relation to various diseases, disorders, states of consciousness, and tasks,
% \DIFadd{Czy many jakieś odnośniki do tego? Np. \url{https://doi.org/10.1098/rsif.2015.1027}, ale Dante tam nie liczył klastrów.} 
this work offers an important consideration towards the robustness of these findings.

In addition to applications to neuroscience, our discussion is relevant in studies of artificial neural networks\cite{DelPapa2017,Landmann2021} where, for the specific learning dynamics of the connectivity matrix, the network self-organizes towards a critical state. In such models, typical criticality indicators are based on avalanche sizes, which are structure-agnostic like the cluster sizes considered in this work.

%TC:ignore % count words wihtou methods, references and appendices
\section{Methods}
\label{sec:methods}

\subsection{Connectivity matrices}
\label{sec:methods_connectomes}

\paragraph{Connectomes of post-stroke brains}
% \label{sec:stroke}
The connectivity matrices that we used in Haimovici model in Sec.~\ref{sec:real_and_artificial}
were a set of 113 individual connectomes from 79 stroke patients and 47 connectomes from 28 control subjects
acquired via Diffusion-Weighted Magnetic Resonance Imaging (DWI)\cite{Corbetta2015,Siegel2016,Rocha2022}.
The patients' connectomes were acquired twice, 3 months ($t_1$) and 12 months ($t_2$) after the stroke.
Each connectome encodes a network of $N=324$ nodes of cortical ROIs, based on Gordon's parcellation\cite{Gordon2016}. 
By convention, the diagonal elements of $W$ are set to zero.

\paragraph{Connectomes in artificial strokes}
For the minimal model of artificial strokes the Hagmann et al.'s connectome\cite{Hagmann2008} served as the base connectivity matrix representing the healthy brain.

\paragraph{Watts-Strogatz networks}
For the Haimovici model results presented in Fig.~\ref{fig:watts-strogatz_stroke_dataset_comparison}\textbf{d}, the empirical connectomes were substituted with Watts-Strogatz small-world networks\cite{watts1998NCollective} of comparable size ($ N=2000$ nodes).
These networks are constructed from a ring of nodes, each symmetrically connected to its nearest neighbors with $k$ edges whose ends are subsequently rewired to random nodes with probability $\pi$.
Following\cite{Zarepour2019,Diaz2021}, to mimic the weight distribution of the human connectome\cite{Hagmann2008}, the link weights were sampled from an exponential distribution $p(w) = \lambda e^{-\lambda w},$ with $\lambda = 12.5$.
The same Greenberg-Hastings\cite{Greenberg1978} dynamics was used as for the Haimovici model,
with $t_\text{max}=10000$, $t_\text{init}= 200$, $r_1 = 0.001$, $r_2 = 0.3$, and the unnormalized symmetric weight matrices $W$.
Node degrees and rewiring probability in WS networks were set to $k=2, \pi=0.5$ to obtain non-critical behavior and to $k=10, \pi=0.5$ to obtain the critical one.

\subsection{Graph connectivity measures}
\label{sec:methods_graph}
% \subsection{Graph notation}
First, we set the notation and some basic definitions for weighted directed graphs, such as connectomes, that we adopt throughout the paper.

For a weighted directed graph $G$ with $N$ nodes, we denote its binary adjacency matrix as $A$ and $a_{ij}$ as its element corresponding to the directed connection from node $j$ to $i$, and analogously the weighted adjacency matrix as $W$ and $w_{ij}$ as the weight of the directed connection.
The analog of degree in weighted directed graphs are in-degree and out-degree strength: $w^{\text{in}}_i = \sum_{j} w_{ij}, w^{\text{out}}_i = \sum_{j} w_{ji}$.
The total strength of the graph is $2 w = \sum_i w^{\text{in}}_i = \sum_i w^{\text{out}}_i = \sum_{i}\sum_{j} w_{ij}$, and the average in-degree strength is $\langle w^{\text{in}} \rangle = \sum_i w^{\text{in}}_i/N$.

In a graph $G$ with the set of nodes $V$, \textit{conductance} of a node subset $S \subset V$ and its complement $\Bar{S}=V\setminus S$ is the quotient\cite{Chung1997}
\begin{equation}
    h_G(S, \Bar{S}) = \frac{\vert \text{cut}(S,\Bar{S})\vert}{\text{min}(\text{vol}(S),\text{vol}(\Bar{S}))}
\end{equation}
of the weighted size $\vert\text{cut}(S,\Bar{S})\vert = \sum_{i\in S,j\in \Bar{S}} w_{ij}+w_{ji}$ of the cut, i.e., the set of all edges connecting $S$ and $\Bar{S}$, and the smaller of the total strengths $\text{vol}(S) = \sum_{i\in S} w^{\text{out}}_i$ summed over all nodes belonging to these sets.
In our study, the connectome was divided into the nodes of the chosen RSN and the rest of the network.

Graph \textit{modularity} $Q$ can be computed for weighted graphs both undirected (Hagmann et al.'s connectome-based networks) and directed graphs (stroke dataset connectomes)\cite{arenas2007NJPSize} %Leicht2008,
using the weighted adjacency matrix $w_{ij}$: 
\begin{equation}
    Q = \frac{1}{2w}\sum_{i, j} \left(w_{ij}- \frac{w^\text{out}_i w^\text{in}_{j}}{2w}\right)\delta_{c_i, c_j},
    % Q = \frac{1}{2m}\sum_{\nu, \omega} \left(W_{\nu,\omega}- \frac{w^\text{out}_\nu w^\text{in}_{\omega}}{2m}\right)\delta_{c_\nu, c_\omega},
\end{equation}
where $c_i$ is the label of the module to which node $i$ belongs, and $\delta_{c_i, c_j}$ is the Kronecker delta symbol of two such labels.
In our work, the modularity was always normalized to the \emph{maximal modularity} of a perfectly mixed network, 
\begin{equation}
    % Q_\text{max} = 1 - \frac{1}{2m}\sum_{\nu,\omega}\frac{k^\text{out}_\nu k^\text{in}_{\omega}}{2m}\delta_{c_\nu, c_\omega}. 
    Q_\text{max} = 1 - \frac{1}{2w}\sum_{i,j}\frac{w^\text{out}_i w^\text{in}_{j}}{2w}\delta_{c_i, c_j}. 
\end{equation}
The results presented for artificially modified connectomes are averages of 20 modification realizations for each value of parameters used.

We used the implementation of conductance and of Louvain modularity optimization algorithm from NetworkX Python package\cite{hagberg2008P7PSCExploring}.

\paragraph{Normalization in Figure~\ref{fig:structural_analysis}}

To compare artificial and real strokes in Fig.~\ref{fig:structural_analysis}, we propose a normalization of quantities defined as
$$A \text{ (norm.)}= (A - A^0)/A^0,$$
where $A^0$ denotes the value of the quantity in the unmodified system (i.e., a system not affected by stroke). Systems whose normalized parameters are close to zero are very similar to the unmodified system, whereas large values of normalized parameters indicate a large deviation from it. 

For real strokes, as the proxy of the unmodified system, we use the control group averages.

\subsection{Haimovici model}
\label{sec:methods_htc}
For simulations of the Haimovici model, we use a publicly available Python code based on the Susceptible-Excited-Refractory model\cite{Damicelli2019}. The simulation runs in time steps, each with the following three transitions performed on all nodes concurrently\cite{Greenberg1978}:
\begin{enumerate}
    \item Active $\to$ Refractory; All nodes activated in the previous time step become dormant.
    \item Refractory $\to$ Inactive; Each dormant node may become inactive with probability $r_2$.
    \item Inactive $\to$ Active; Inactive nodes become active in two ways: a) due to spontaneous activation with probability $r_1$ and b) due to neighboring active nodes that satisfy the threshold criterion $\sum_{j \ \text{active}} w_{ij} > \mathcal{T}$.
\end{enumerate}

Numerical simulations are performed for $t_\text{max}=10000$ time steps for $30$ threshold values $\mathcal{T}$, covering the critical value $\mathcal{T}_c$. Each simulation starts in a random state with $1\%$ of nodes active and the rest inactive. To account for this initial randomness, the first $t_\text{init}=200$ time steps in all simulations are discarded from further analysis, leaving $t_\text{sim}=t_\text{max}-t_\text{init}$ simulation steps. The result, for each $\mathcal{T}$, is a $N\times t_\text{sim}$ data matrix 
% $X_{it}$
whose entries take the values $0$ (inactive nodes), $1$ (active nodes) and $2$ (refractory nodes). Following the literature\cite{Haimovici2013, Rocha2018}, after finishing the simulation, we apply a preprocessing step of conflating the inactive and refractory states by substituting $2 \to 0$ in the data matrix.
% $X$.

The states of nodes $s_i(t)$ at time $t$ are used to calculate the total activity $A(t) = \sum_{i=1}^N s_i(t)$. The average activity $\langle A \rangle$ and the deviation of the total activity $\sigma_A$ are defined as follows:
\begin{equation}
    \langle A\rangle = \frac{1}{t_\text{sim}}\sum_{t=1}^{t_\text{sim}} A(t), \quad \sigma_A = \sqrt{\frac{1}{t_\text{sim}} \sum_{t=1}^{t_\text{sim}}(A(t) - \langle A\rangle)^2}.
\end{equation}

An approximate critical value of the threshold in the Haimovici model can be computed using the mean-field approach\cite{Rocha2018}:
\begin{equation}
    % \mathcal{T}_c = \langle W \rangle \frac{r_2}{1+2r_2},
    \mathcal{T}_c = \langle w^{\text{in}} \rangle \frac{r_2}{1+2r_2},
\end{equation}
where $r_2$ is the probability of transition from the refractory state to the inactive state.
In all our calculations, following previous studies\cite{Rocha2018}, we keep $r_2={r_1}^{0.2}$, where $r_1$ is the probability of spontaneous activation set to $r_1=2/N$. For Hagmann et al.'s connectome, the mean-field critical threshold value is $\mathcal{T}_c=0.08$. This theoretical value agrees quite well with the results of our numerical simulations.

\paragraph{Modifications to the Haimovici model} % introduced in\cite{Rocha2022}
As a way of implementing the homeostatic plasticity principle in the Haimovici model, the network excitability was balanced by normalizing the incoming node’s excitatory input in the structural connectivity matrix\cite{Rocha2018}:
% $$\tilde{W}_{ij} = \frac{W_{ij}}{\sum_j W_{ij}}.$$ 
% $$\tilde{w}_{ij} = \frac{w_{ij}}{w^{in}_{i}}.$$ 
$$\tilde{w}_{ij} = w_{ij}/w^{in}_{i}.$$ 
Such equalization minimizes the variability of activity and of the position of the critical point between individual connectivity matrices.
It is noteworthy that the control parameter also becomes rescaled $\mathcal{T}\rightarrow\tilde{\mathcal{T}}$, so that the critical point moves to an empirical value $\tilde{\mathcal{T}}_c \approx 0.15$. 
The model exhibits a similar behavior of the cluster sizes as the Haimovici model, where at the critical value of the threshold the size of the second largest cluster peaks.
Similarly to the Haimovici model, in numerical simulations, we use $r_1=2/N$ and $r_2={r_1}^{0.2}$ and simulate the activity of the system for $t_\text{max}=10000$ time steps.

The normalized connectivity matrices $\tilde{W}$ from the stroke data set used in\cite{Rocha2022} are available publicly\cite{rocha2022zenodo}.

\subsection{Clusters}
\label{sec:methods_clusters}
We define clusters as maximal sets of nodes sharing the same type of activity ($\pm 1$ in the Ising model and active in the Haimovici model) connected following the adjacency matrix. The size of the cluster is the total number of participating nodes. Among all cluster sizes, we focus on the two largest cluster sizes $S_1$ and $S_2$, which are standard order parameters in percolation theory\cite{stauffer1994}. We measured the sizes averaged over the simulation time $t_\text{sim}$:
\begin{align}
    S_i = \frac{1}{t_\text{sim}} \sum_{t=1}^{t_\text{sim}} S_i(t), \qquad i = 1,2,
\end{align}
where $S_i(t)$ is the size of the momentary $i$-th largest cluster found at time $t$.

\subsection{Ising model}
\label{sec:methods_ising}

The Ising model on a lattice with the adjacency matrix $A$ is defined in terms of the energy function:
\begin{equation}\label{eqn:e}
    E = - J \sum_{i,j=1}^N a_{ij} s_i s_j,
\end{equation}
where the spin variables $s_i$ take values $\pm 1$ and $J$ is the coupling parameter.
% The square lattice we use for our model has dimensions $100 \times 100$ with nonperiodic boundary conditions. 
% The size of the system $N = 10000$ is the total number of lattice sites. 
The probability of a particular spin configuration $s=(s_1,s_2,\ldots,s_N)$ depends on temperature $T$ according to the Boltzmann distribution
$$ P(s)\sim e^{-E(s)/T}$$
and the approximate value of the critical temperature is $T_c = 2.27$, as shown in a classic paper by Kramers and Wannier\cite{KramersWannier1941}.

\begin{figure}[h]
    \centering
    \includegraphics[scale=.5]{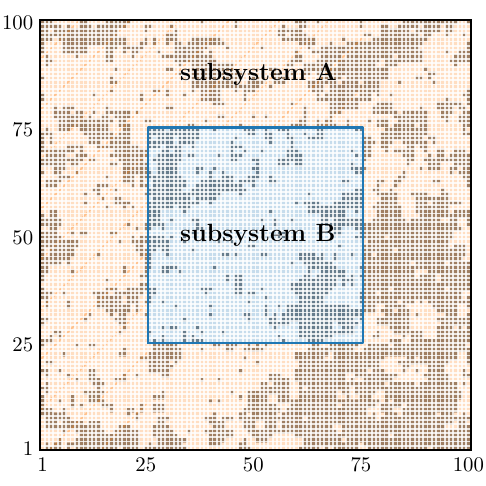}
    \caption{\textbf{Unequal Ising subsystems.} Division of the Ising model lattice into unequal sizes $N_A > N_B$.}
    \label{fig:ising_patches}
\end{figure}

\paragraph{Details on numerical simulations}
The coupling parameter was set to unity $J=1$ and a square lattice of dimensions $100 \times 100$ with nonperiodic boundary conditions was used.
The size of the system $N = 10000$ was the total number of lattice sites.
In the case $N_\text{A}=N_\text{B}$, the system was divided into two rectangles $100 \times 50$.
In the case $N_\text{A} > N_\text{B}$, subsystem B was defined as a square patch in the center of the lattice and subsystem A as its complement (see Fig.~\ref{fig:ising_patches}), and modified the connections accordingly.

We performed numerical simulations using Monte Carlo importance sampling\cite{LandauBinder2014}. Spins in the initial conditions were set to $-1$ with probability $0.75$ and to $1$ otherwise. Each time step (or sweep) comprised $N$ spin-flips, which in turn followed the Metropolis approach\cite{Metropolis1953}. First, the $i$-th node was flipped $s_i \to - s_i$ and the resulting change in energy was computed $\Delta E = E(\text{spin-flip}) - E_0$. The new configuration was accepted with probability $\min ( e^{-\frac{1}{T} \Delta E}, 1 )$. Simulations were run for $30$ equally spaced temperature values between $T=0.01$ and $T=4.5$, each simulation continuing for $t_\text{max} = 5000$ sweeps with the $t_\text{init} = 200$ initial sweeps discarded.
The resulting data matrix for a single temperature had dimensions $N \times (t_\text{max} - t_\text{init})$.

\subsection{Differences between Ising and Haimovici models}
\label{sec:methods_comparison}

Here, we describe differences between the Ising and the Haimovici models which are important for understanding the comparison between the models but are largely immaterial to our argument.
The Haimovici model has starkly different dynamics than the Ising model. In the latter case, each spin, whether pointing up or down, always belongs to some cluster, unless it is surrounded by four spins of the opposite orientation. This results in large domains such that, in the low $T$ regime, 
the largest and second-largest clusters can cover virtually the entire subsystems A and B, respectively (see Fig.~\ref{fig:ising_clusters}\textbf{b}). In contrast, in the Haimovici model, it is only the clusters of \emph{active} nodes that are considered. An activated node becomes refractory at the very next time step and then waits to become inactive again (with the probability of becoming inactive at a given step $r_2\approx 0.29$). This burst-like cluster formation results in the average sizes of the largest and second-largest clusters attaining much lower maximal values ($S_1 \approx 0.17N$ for $\mathcal{T}$ near $0$; $S_2 \approx 0.006N$ at its peak) than in the case of the Ising model and rapidly decreasing to zero for increasing values of $\mathcal{T} > \mathcal{T}_c$. The unmodified networks in both models are also significantly different: the Haimovici model uses Hagmann et al.'s connectome, which belongs to the small-world class\cite{Zarepour2019}, whereas the Ising model uses a regular two-dimensional square lattice. 

When comparing both models, one should be mindful that the role of the threshold parameter $\mathcal{T}$ is inverse to that of the temperature $T$ in the Ising model, i.e., small values of $\mathcal{T}$ result in supercritical states with high total activity, and large values of $\mathcal{T}$ result in subcritical states with low stochastically induced activity.

\section{Supplementary Information}

\subsection{Investigation of various measures of criticality}
\label{app:otherq}

We calculate an extended list of criticality measures for the empirical connectomes of stroke patients and for the critical and non-critical human-connectome-based Watts-Strogatz networks. This Supplementary Information serves as an extension to Fig.~\ref{fig:watts-strogatz_stroke_dataset_comparison}
where the three most relevant quantities were shown. The results of this extended study are reported in Supplementary Fig.~\ref{fig:supp_crit}. 

\begin{figure}[hbpt]
    \centering
    \includegraphics[width=0.9\linewidth]{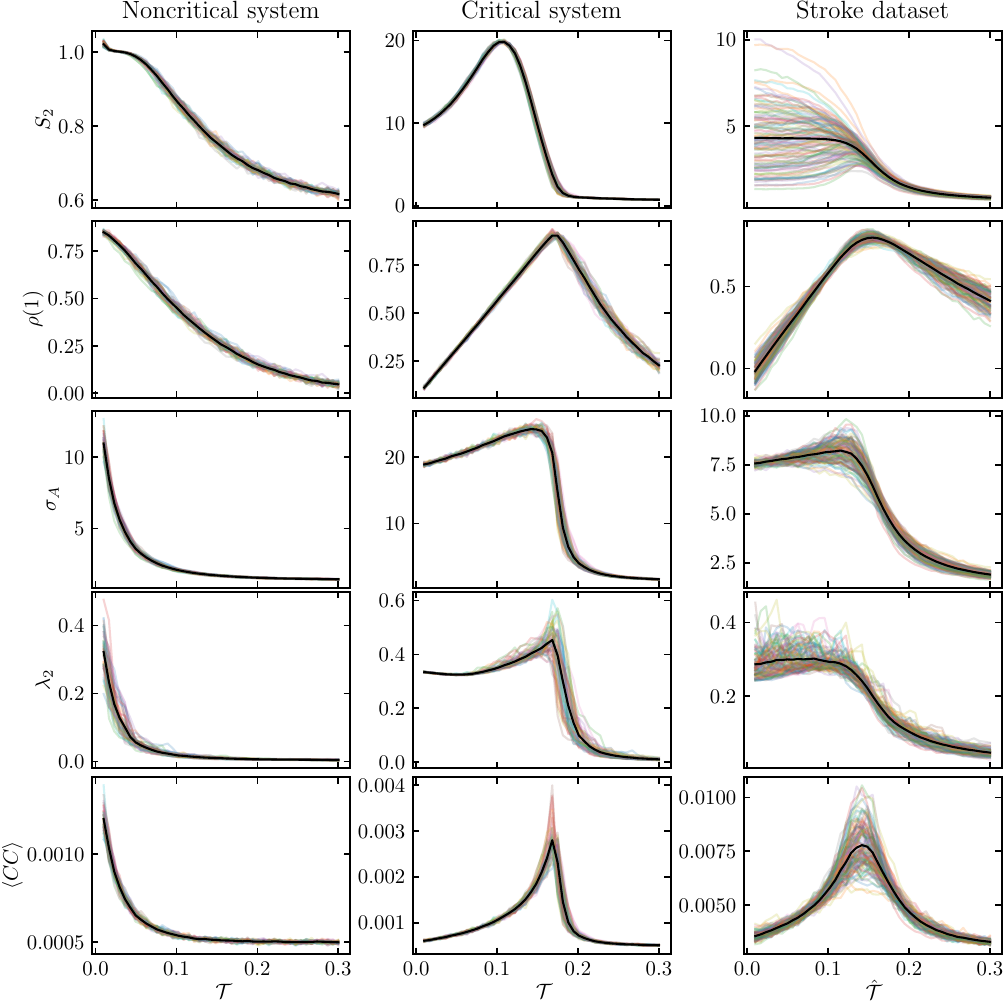}
    \caption{\textbf{Criticality measures in the Haimovici model.} \textbf{Left and middle:} noncritical and critical connectome-based Watts-Strogatz networks. \textbf{Right:} stroke dataset connectomes. The measures include: $S_2$, the size of the second largest cluster of activity; $\rho(1)$, the first coefficient of the autocorrelation function; $\sigma_A$, the standard deviation of the total activity; $\lambda_2$, the second largest eigenvalue of the node activity cross-correlation matrix; $\langle CC \rangle$, the average of cross-correlation matrix elements. }
    \label{fig:supp_crit}
\end{figure}

\begin{figure}[hbpt]
    \centering
    \includegraphics[width=0.9\textwidth]{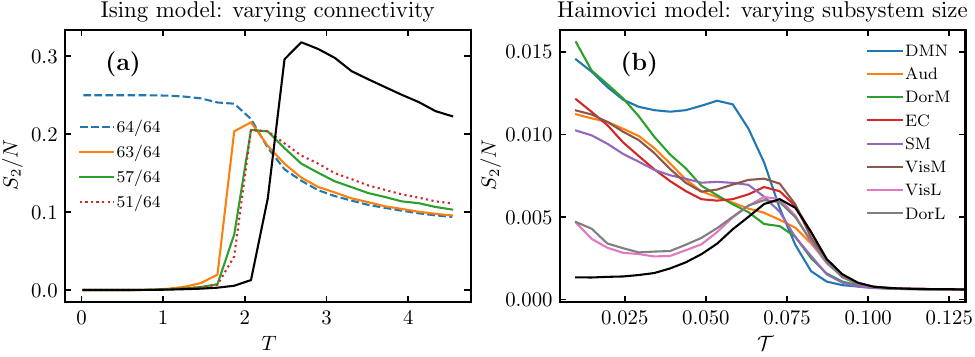}
    \caption{Dependence of the size of the second largest cluster, $S_2$, on: \textbf{(a)} temperature for various fractions of removed connections between the subsystems in the Ising model on a lattice of size $32\times 32$ with the smaller subsystem defined as a $16\times 16$ patch,
    and \textbf{(b)} the threshold parameter for fully disconnected RSN of varying sizes and locations in the Haimovici model. The legend is ordered from the smallest resting-state network \emph{DorL} ($46$ nodes) to the largest \emph{DMN} ($128$ nodes). Black lines denote cluster sizes $S_2$ for unmodified systems. These plots supplement Fig.~\ref{fig:ising_htc_comparison}
    by inspecting other ways of modifying the systems than those described in the main text.
    }
    \label{fig:supp_ising_htc_comparison}
\end{figure}

\subsection{Other modifications of the models}
\label{sec:suppA}

In this section, we apply different modifications to the Ising and Haimovici models to demonstrate the robustness of the main result presented in Fig.~\ref{fig:ising_htc_comparison}.
We swap the modifications in a complementary way. We gradually disconnect subsystem B in the Ising model (in the main text, we varied the subsystem size), and we vary the subsystem size in the Haimovici model by disconnecting different single RSNs (in the main text, we gradually disconnected only the auditory RSN). Both results are shown in Supplementary Fig.~\ref{fig:supp_ising_htc_comparison}. In this study, the Ising model runs on a $32\times 32$ lattice and subsystem B is a patch of size $16 \times 16$.

In the Ising case, Supplementary Fig.~\ref{fig:supp_ising_htc_comparison}\textbf{a}, we find a family of curves forming a degenerate comb with its upper non-critical branch consisting of a single line corresponding to all connections removed. The remaining modifications cover the lower branch of the comb and exhibit indicators of criticality. This behavior is characteristic of the model in which even a single connection between the subsystems is sufficient to connect the clusters and make the behavior qualitatively the same as in an unmodified system.

Disconnecting different RSNs serves as a proxy for varying the size of the subsystem. In this case, the resulting family of curves does form a comb-like pattern, albeit somewhat perturbed; this is not a surprise, as different brain substructures are not rescaled copies of each other. Still, the curves are size-ordered to a certain degree. The most prominent examples of partitions that preserve the critical peak (such as \emph{DorL} and \emph{VisL}) are the smallest RSNs, and those that appear non-critical (such as \emph{DMN}) are the largest ones. This is in agreement with the Ising case with varying subsystem size, as seen in Fig.~\ref{fig:ising_htc_comparison}\textbf{a}.
The competition between $S_2^\text{A}$ and $S_1^\text{B}$ and the resulting second largest cluster of the entire system when individual RSNs are fully disconnected is additionally depicted in Supplementary Fig.~\ref{fig:supp_rsn_cluster_hierarchy}.

\begin{figure}[htbp!]
    \centering
    \includegraphics[width=0.9\textwidth]{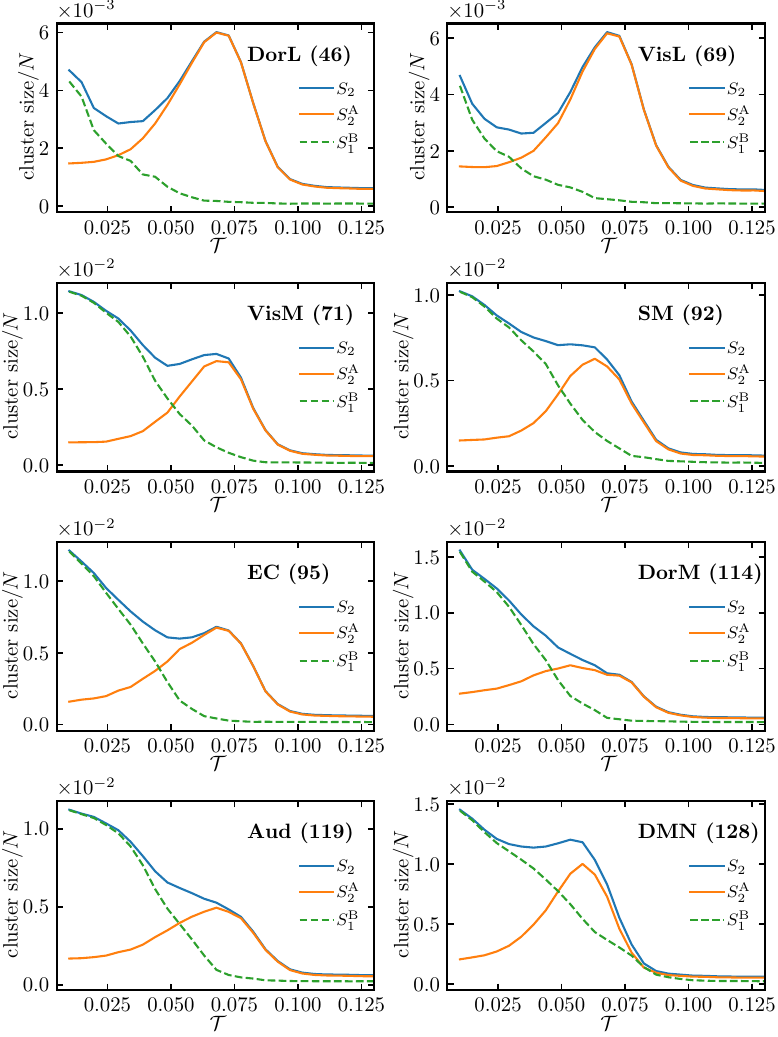}
    \caption{Cluster sizes in the Haimovici model for the Hagmann et al.'s connectome with various RSNs fully disconnected from the rest of the network. Each panel shows the second largest cluster of the entire system, $S_2$ (blue line), the second largest cluster of the larger subsystem, $S_2^\text{A}$ (orange line), and the largest cluster of the disconnected RSN (abbreviation and size give in the panel), $S_2^\text{B}$ (dashed green line). In all the cases, at some value of the threshold parameter $\mathcal{T}$, the order of sizes changes, impacting the size of the second largest entire-system cluster. 
    Depending on the case, the peak in $S_2$ may persist (e.g., \emph{DorL}, \emph{VisL}) or be absent (e.g., \emph{DorM}, \emph{Aud}, and \emph{DMN}).}
    \label{fig:supp_rsn_cluster_hierarchy}
\end{figure}

\newpage
\bibliography{main}

\section*{Acknowledgements}

We would like to thank Dante R. Chialvo and Maciej A. Nowak for fruitful discussions and critical advice throughout the duration of this project. The authors acknowledge helpful conversations with Rodrigo P. Rocha about the usage of stroke data. This work was carried out within the research project ‘‘Bio-inspired artificial neural networks’’
(grant no. POIR.04.04.00-00-14DE/18-00) within the Team-Net
program of the Foundation for Polish Science co-financed by the
European Union under the European Regional Development Fund.
The research for this publication has been supported by a grant from the Priority Research Area DigiWorld under the Strategic Programme Excellence Initiative at Jagiellonian University.
JJ acknowledges the support of the Faculty of Physics, Astronomy and Applied Computer Science of the Jagiellonian University through grant no. LM/3/JJ.

\section*{Author contributions statement}
JJ, ZD, and JG conducted the experiments, and JO and PO analyzed the results.  All authors reviewed the manuscript. 

\section*{Additional information}
\textbf{Competing interests} The authors declare that they have no conflict of interest. \newline
\textbf{Materials \& Correspondence} Correspondence should be addressed to Jeremi K. Ochab (jeremi.ochab@uj.edu.pl).

\section*{Code availability}
Scripts used in the simulations and data analysis are publicly available \url{https://github.com/grelade/critical-stroke}.

\section*{Data availability}
The connectomes used in this work come from\cite{Rocha2022} and are available from Zenodo repository\cite{rocha2022zenodo}. 
They are derived from the raw neuroimaging data from\cite{Corbetta2015,Siegel2016}, which are publicly available at cnda.wustl.edu and require controlled access as they contain sensitive patients’ data.

% %TC:endignore

\end{document}